\documentclass[
    ,final            
  ]
  {aipproc}

\layoutstyle{6x9}

\usepackage{graphicx}
\usepackage{subfigure}
\usepackage{psfrag}
\usepackage{amsbsy, amssymb}

\newcommand*{\myalign}[2]{\multicolumn{1}{#1}{#2}}

\def\reff@jnl#1{{\rm#1\/}}
\def\aj{\reff@jnl{AJ}}                 
\def\araa{\reff@jnl{ARA\&A}}           
\def\apj{\reff@jnl{ApJ}}               
\def\apjl{\reff@jnl{ApJ}}              
\def\apjs{\reff@jnl{ApJS}}             
\def\ao{\reff@jnl{Appl.Optics}}        
\def\apss{\reff@jnl{Ap\&SS}}           
\def\aap{\reff@jnl{A\&A}}              
\def\aapr{\reff@jnl{A\&A~Rev.}}        
\def\aaps{\reff@jnl{A\&AS}}            
\def\azh{\reff@jnl{AZh}}               
\def\baas{\reff@jnl{BAAS}}             
\def\jrasc{\reff@jnl{JRASC}}           
\def\memras{\reff@jnl{MmRAS}}          
\def\mnras{\reff@jnl{MNRAS}}           
\def\jcap{\reff@jnl{JCAP}}             
\def\pra{\reff@jnl{Phys.Rev.A}}        
\def\prb{\reff@jnl{Phys.Rev.B}}        
\def\prc{\reff@jnl{Phys.Rev.C}}        
\def\prd{\reff@jnl{Phys.Rev.D}}        
\def\prl{\reff@jnl{Phys.Rev.Lett}}     
\def\pasp{\reff@jnl{PASP}}             
\def\pasj{\reff@jnl{PASJ}}             
\def\qjras{\reff@jnl{QJRAS}}           
\def\skytel{\reff@jnl{S\&T}}           
\def\solphys{\reff@jnl{Solar~Phys.}}   
\def\sovast{\reff@jnl{Soviet~Ast.}}    
\def\ssr{\reff@jnl{Space~Sci.Rev.}}    
\def\zap{\reff@jnl{ZAp}}               
\def\nat{\reff@jnl{Nature}}            

\begin{document}

\title{Exploring Multi-Modal Distributions with Nested Sampling}


\author{Farhan Feroz}{
  address={Cavendish Laboratory, Astrophysics Group, Cambridge CB3 0HE}
}

\author{John Skilling}{
  address={Maximum Entropy Data Consultants, Kenmare, Ireland}
}

\begin{abstract}
In performing a Bayesian analysis, two difficult problems often emerge. First, in estimating the parameters of some model for the data, the resulting posterior distribution may be multi-modal or exhibit pronounced (curving) degeneracies. Secondly, in selecting between a set of competing models, calculation of the Bayesian evidence for each model is computationally expensive using existing methods such as thermodynamic integration. Nested Sampling is a Monte Carlo method targeted at the efficient calculation of the evidence, but also produces posterior inferences as a by-product and therefore provides means to carry out parameter estimation as well as model selection. The main challenge in implementing Nested Sampling is to sample from a constrained probability distribution. One possible solution to this problem is provided by the Galilean Monte Carlo (GMC) algorithm. We show results of applying Nested Sampling with GMC to some problems which have proven very difficult for standard Markov Chain Monte Carlo (MCMC) and down-hill methods, due to the presence of large number of local minima and/or pronounced (curving) degeneracies between the parameters. We also discuss the use of Nested Sampling with GMC in Bayesian object detection problems, which are inherently multi-modal and require the evaluation of Bayesian evidence for distinguishing between true and spurious detections.
\end{abstract}

\maketitle


\section{Introduction}

Bayesian inference provides a consistent approach to the estimation of a set of parameters $\boldsymbol \Theta$ in a model (or hypothesis) $H$ for the data $\boldsymbol D$. Bayes' theorem states that
\begin{equation} 
\Pr(\boldsymbol \Theta|\boldsymbol D, H) = \frac{\Pr(\boldsymbol D|\,\boldsymbol \Theta,H)\Pr(\boldsymbol \Theta|H)}{\Pr(\boldsymbol D|H)},
\end{equation}
where $\Pr(\boldsymbol \Theta|\boldsymbol D, H) \equiv P(\boldsymbol \Theta|\boldsymbol D)$ is the posterior probability distribution of the parameters, $\Pr(\boldsymbol D|\boldsymbol \Theta, H) \equiv \mathcal{L}(\boldsymbol \Theta)$ is the likelihood, $\Pr(\boldsymbol \Theta|H) \equiv \pi(\boldsymbol \Theta)$ is the prior, and $\Pr(\boldsymbol D|H) \equiv \mathcal{Z}$ is the Bayesian evidence, which is the factor required to normalize the posterior over $\boldsymbol \Theta$ and is given by:
\begin{equation}
\mathcal{Z} = \int{\mathcal{L}(\boldsymbol \Theta)\pi(\boldsymbol \Theta)}d^n\boldsymbol \Theta,
\label{eq:Z}
\end{equation} 
where $n$ is the dimensionality of the parameter space. Bayesian evidence being independent of the parameters, can be ignored in parameter estimation problems and inferences can be obtained by taking samples from the (unnormalized) posterior distribution using standard MCMC methods.

Model selection between two competing models $H_{0}$ and $H_{1}$ can be done by comparing their respective posterior probabilities given the observed data-set $\boldsymbol D$, as follows
\begin{equation}
R = \frac{\Pr(H_{1}|\boldsymbol D)}{\Pr(H_{0}|\boldsymbol D)}
  = \frac{\Pr(\boldsymbol D|H_{1})\Pr(H_{1})}{\Pr(\boldsymbol D| H_{0})\Pr(H_{0})}
  = \frac{\mathcal{Z}_1}{\mathcal{Z}_0} \frac{\Pr(H_{1})}{\Pr(H_{0})},
\label{eq:R}
\end{equation}
where $\Pr(H_{1})/\Pr(H_{0})$ is the prior probability ratio for the two models, which can often be set to unity in situations where there is not a prior reason for prefering one model over the other, but occasionally requires further consideration. It can be seen from Eq.~(\ref{eq:R}) that the Bayesian evidence plays a central role in Bayesian model selection.

As the average of the likelihood over the prior, the evidence is larger for a model if more of its parameter space is likely and smaller for a model with large areas in its parameter space having low likelihood values, even if the likelihood function is very highly peaked. Thus, the evidence automatically implements Occam's razor.

\section{Nested Sampling}

Nested sampling \cite{skilling04, sivia} is a Monte Carlo technique to estimate the Bayesian evidence by transforming the multi-dimensional evidence integral into a one-dimensional integral. This is accomplished by defining the prior volume $X$ as $dX = \pi(\boldsymbol \Theta)d^n \boldsymbol \Theta$, so that
\begin{equation}
X(\lambda) = \int_{\mathcal{L}\left(\boldsymbol \Theta\right) > \lambda} \pi(\boldsymbol \Theta) d^n\boldsymbol \Theta,
\label{eq:Xdef}
\end{equation}
where the integral extends over the region(s) of parameter space contained within the iso-likelihood contour $\mathcal{L}(\boldsymbol \Theta) = \lambda$. The evidence integral, Eq.~(\ref{eq:Z}), can then be written as:
\begin{equation}
\mathcal{Z}=\int_{0}^{1}{\mathcal{L}(X)}dX,
\label{eq:nested}
\end{equation}
where $\mathcal{L}(X)$, is a monotonically decreasing function of $X$. Thus, by evaluating the likelihoods $\mathcal{L}_{i}=\mathcal{L}(X_{i})$, where $X_{i}$ is a sequence of decreasing values,
\begin{equation}
0<X_{M}<\cdots <X_{2}<X_{1}< X_{0}=1.
\end{equation}
Evidence can then be approximated numerically using standard quadrature methods as follows:
\begin{equation}
\mathcal{Z}={\textstyle {\displaystyle \sum_{i=1}^{M}}\mathcal{L}_{i}w_{i}},
\label{eq:NS_sum}
\end{equation}
where the weights $w_{i}$ for the simple trapezium rule are given by $w_{i}=\frac{1}{2}(X_{i-1}-X_{i+1})$.

The summation in Eq.~(\ref{eq:NS_sum}) is performed as follows. First $N$ `live' points are drawn uniformly from the prior distribution $\pi(\boldsymbol \Theta)$ and initial prior volume $X_{0}$ is set to unity. At each subsequent iteration $i$, the point with lowest likelihood value $\mathcal{L}_{i}$ is removed from the live point set and replaced by another point drawn uniformly from the prior distribution with the condition that its likelihood is higher than $\mathcal{L}_{i}$. This results in the new point being drawn uniformly from the prior volume contained within the iso-likelihood contour defined by $\mathcal{L}_{i}$. The prior volume contained within this region at $i^{\rm th}$ iteration, is a random variable given by $X_{i} = t_{i} X_{i-1}$, where $t_{i}$ follows the distribution $\Pr(t) = Nt^{N-1}$ (i.e., the probability distribution for the largest of $N$ samples drawn uniformly from the interval $[0,1]$). This process is repeated, until the entire prior volume has been traversed. The algorithm thus travels through nested shells of likelihood as the prior volume is reduced. The mean and standard deviation of $\log t$, which dominates the geometrical exploration, are:
\begin{equation}
E[\log t] = -1/N, \quad \sigma[\log t] = 1/N.
\end{equation}
Since each value of $\log t$ is independent, after $i$ iterations the prior volume will shrink down such that $\log X_{i} \approx -(i \sqrt{i})/N$. Thus, one takes $X_{i} = \exp(-i/N)$.

\section{Galilean Monte Carlo}

The main challenge in implementing a nested sampling algorithm is to draw samples from the prior distribution with the constraint $\mathcal{L}> \mathcal{L}_i$, where $\mathcal{L}_i$ is lowest likelihood value among all live points at each iteration $i$. One widely used algorithm to approach this problem in astrophysics is \texttt{MultiNest} \cite{feroz08, multinest} which is based on an ellipsoidal rejection sampling scheme. At each iteration $i$, the full set of $N$ live points is enclosed within a set of (possibly overlapping) ellipsoids and a new point is then drawn uniformly from the region enclosed by these ellipsoids. However, this approach becomes inefficient in high dimensional ($n \gtrsim 100$) problems.

An alternative way to draw a point from this constrained distribution is by using MCMC, however it could be very inefficient in problems exhibiting degeneracies between the parameters. Since we need to draw a point uniformly from the region where $\mathcal{L} > \mathcal{L}_i$ and we already have $N$ `live' points distributed uniformly inside this region, we could start a Markov Chain at one of these `live' points with initial velocity $\boldsymbol v$ and reflect off the boundary of this region where $\mathcal{L} = \mathcal{L}_i$ whenever we encounter it. The problem however is that the location of boundary is not known. Galilean Monte Carlo \cite{2012AIPC.1443..145S} addresses precisely this problem.

At each iteration $i$, GMC proceeds by picking a `live' point with coordinates $\boldsymbol x_1$ at random and gives it initial velocity $\boldsymbol v$. A new point $\boldsymbol x_2 = \boldsymbol x_1 + \boldsymbol v$ is then proposed which is accepted if $\mathcal{L}(\boldsymbol x_2) > \mathcal{L}_i$, otherwise a third point $\boldsymbol x_3$ is proposed by reflecting off $\boldsymbol x_2$ i.e. $\boldsymbol x_3 = \boldsymbol x_2 + \boldsymbol v^{\prime}$ where $\boldsymbol v^{\prime} = \boldsymbol v - 2 \boldsymbol n (\boldsymbol n . \boldsymbol v)$ and $\boldsymbol n$ is a unit vector perpendicular to $\nabla \mathcal{L}$ at $\boldsymbol x_2$. If $\mathcal{L}(\boldsymbol x_3) > \mathcal{L}_i$, $\boldsymbol x_3$ is accepted otherwise the trajectory from $\boldsymbol x_1$ is reversed by giving it velocity $\boldsymbol -v$. These moves are repeated for $k$ steps resulting in total path length of $k \boldsymbol v$.

\section{Applications}

In this section we show the results of applying Nested Sampling with the GMC algorithm to several multi-modal toy problems which have proven to be very challenging for MCMC algorithms as they tend to get stuck in isolated modes and have very low efficiencies due to the presence of degeneracies between the parameters. We refer the reader to \cite{DBLP:journals/amc/LiPAZG09} for an example of MCMC based algorithms applied to similar problems.

In order to analyse these problems with GMC, we used $1000$ live points and set the log-likelihood $\log\mathcal{L}(\boldsymbol \Theta) = -f(\boldsymbol \Theta)$, where $\boldsymbol \Theta = (\theta_1,\theta_2,\cdots,\theta_{\rm n})$ is the parameter vector, $n$ is the dimensionality of the problem and $f(\boldsymbol \Theta)$ is the mathematical description of the toy problem.

\subsection{Himmelblau's Function}

\begin{figure}
\includegraphics[width=0.5\columnwidth]{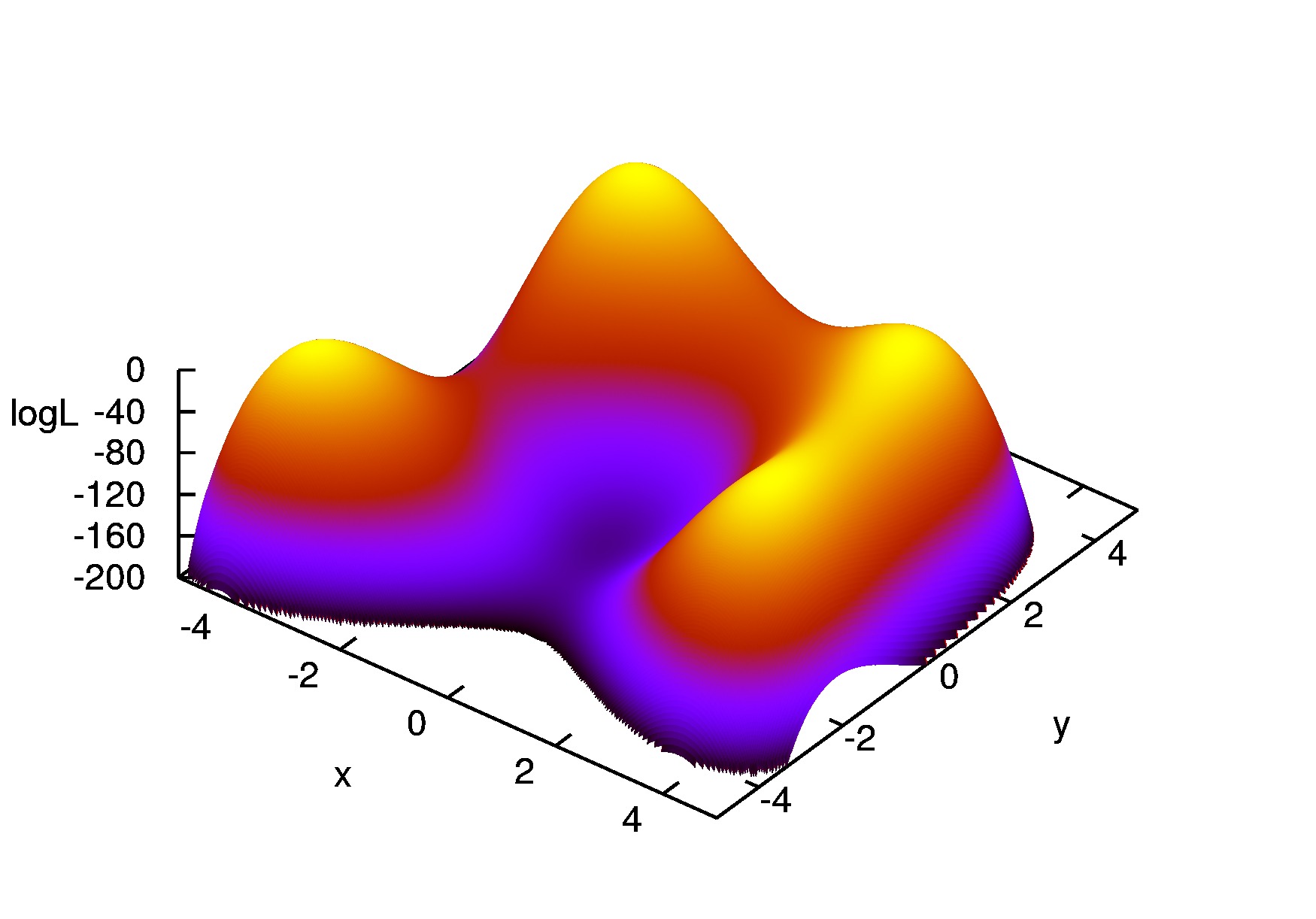}
\includegraphics[width=0.5\columnwidth]{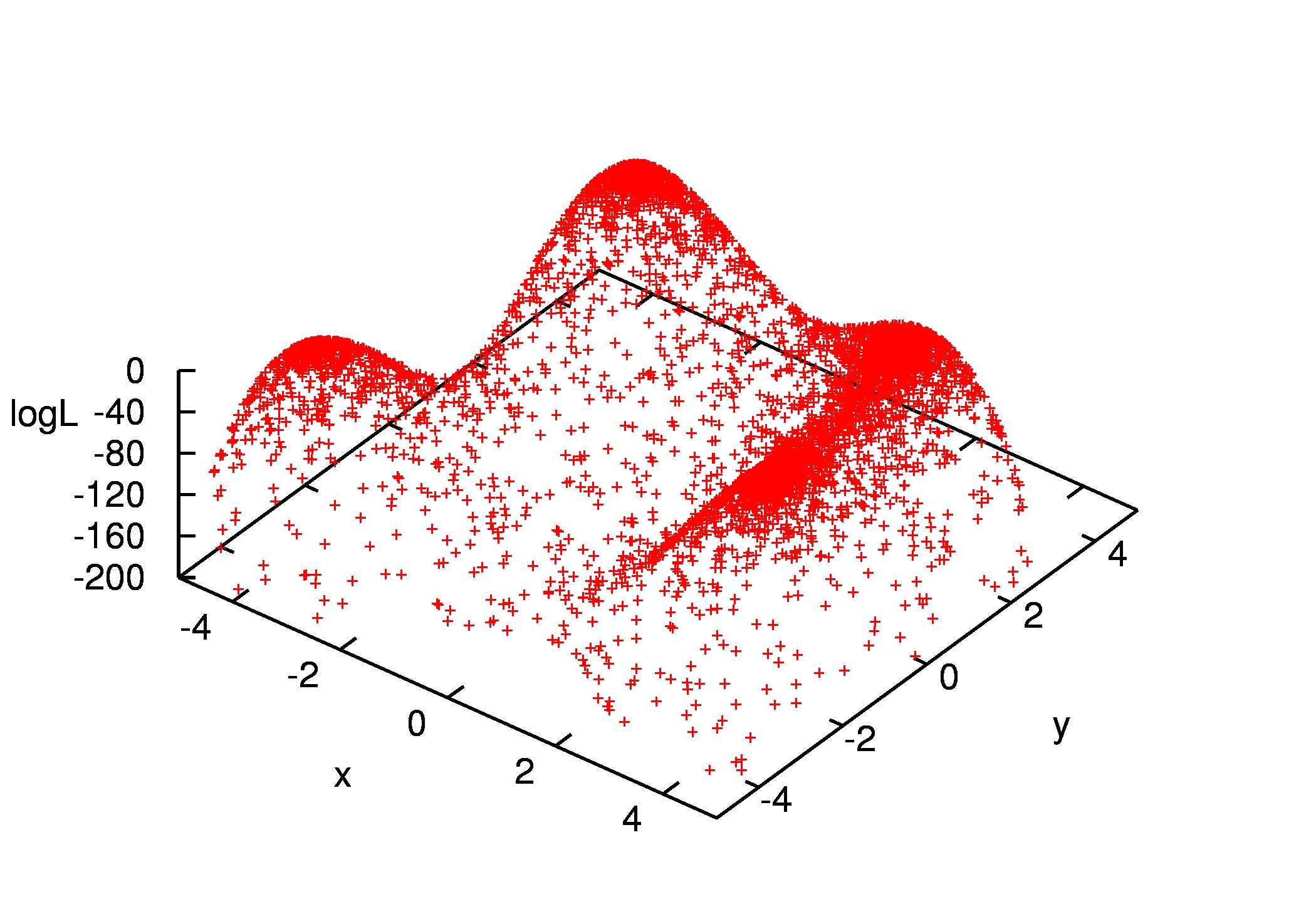}
\caption{Left panel: Himmelblau's function with the z-axis being $\log\mathcal{L}(\boldsymbol \Theta)$. Right panel: Samples obtained by running GMC algorithm on Himmelblau's function.}
\label{fig:himmelblau}
\end{figure}

Himmelblau's function is a 2D function defined as follows:
\begin{equation} 
f(x,y) = (x^2 + y - 11)^2 + (x + y^2 - 7)^2.
\end{equation}
It has four identical local minima at $(3,2)$, $(-2.81, 3.13)$, $(-3.78,-3.28)$ and $(3.58,-1.85)$ where $f(x,y) = 0$. Fig.~\ref{fig:himmelblau} (left panel) shows a plot of the Himmelblau's function with the z-axis being $\log\mathcal{L}(\boldsymbol \Theta) = -f(\boldsymbol \Theta)$. GMC algorithm was run on this problem by assuming uniform priors $\mathcal{U}(-5, 5)$ on both $x$ and $y$. The algorithm took $120,939$ likelihood evaluations and the resultant samples are plotted in Fig.~\ref{fig:himmelblau} (right panel).

\subsection{Eggbox Function}

\begin{figure}
\includegraphics[width=0.5\columnwidth]{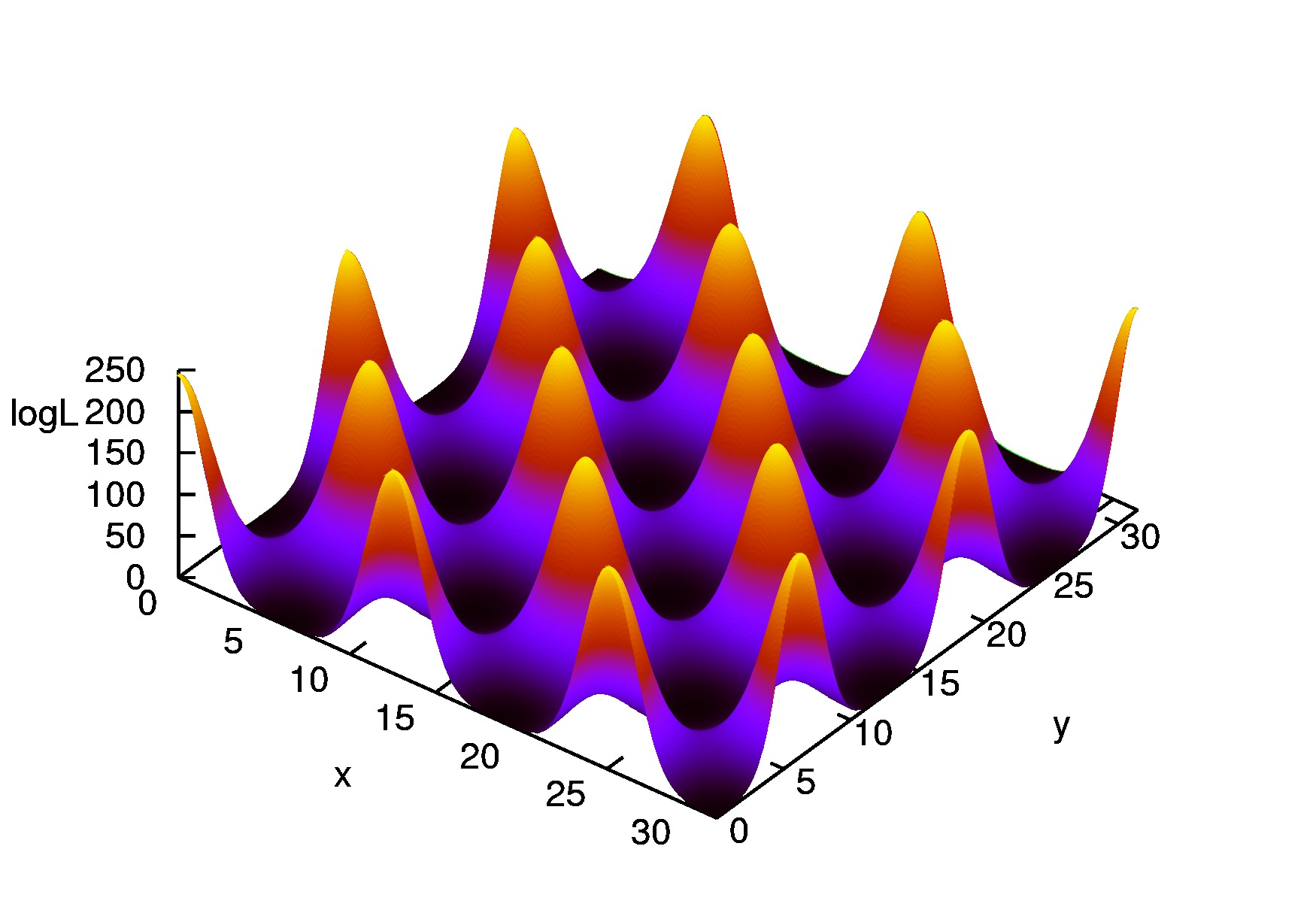}
\includegraphics[width=0.5\columnwidth]{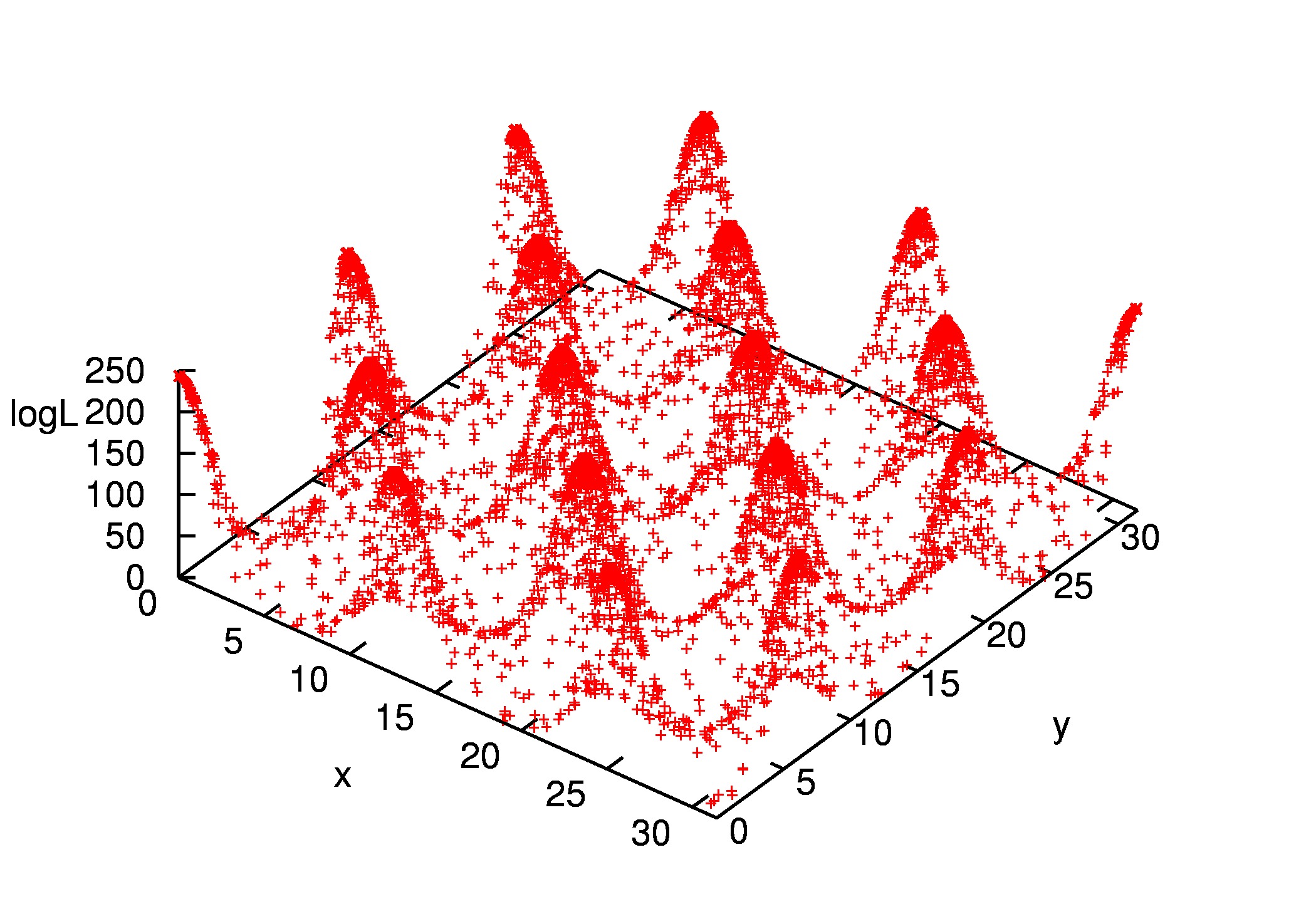}
\caption{Left panel: 2D Eggbox function with the z-axis being $\log\mathcal{L}(\boldsymbol \Theta)$. Right panel: Samples obtained
by running GMC algorithm on 2D Eggbox function.}
\label{fig:eggbox}
\end{figure}

Eggbox function is defined as follows:
\begin{equation} 
f(\boldsymbol \Theta) = - \left[ 2 + \prod_{\rm i}^{n} \cos\left( \frac{\theta_{\rm i}}{2} \right) \right]^5,
\end{equation}
where $\theta_{\rm i} \in [0, 10\pi]$. Fig.~\ref{fig:eggbox} (left panel) shows a plot of the 2D Eggbox function with the z-axis being $\log\mathcal{L}(\boldsymbol \Theta) = -f(\boldsymbol \Theta)$. GMC algorithm was run on this problem by assuming uniform priors $\mathcal{U}(0, 10\pi)$ on parameters $\theta_{\rm i}$. The algorithm took $205,534$ likelihood evaluations and the resultant samples are plotted in Fig.~\ref{fig:eggbox} (right panel).

\subsection{Rastrigin Function}

\begin{figure}
\includegraphics[width=0.5\columnwidth]{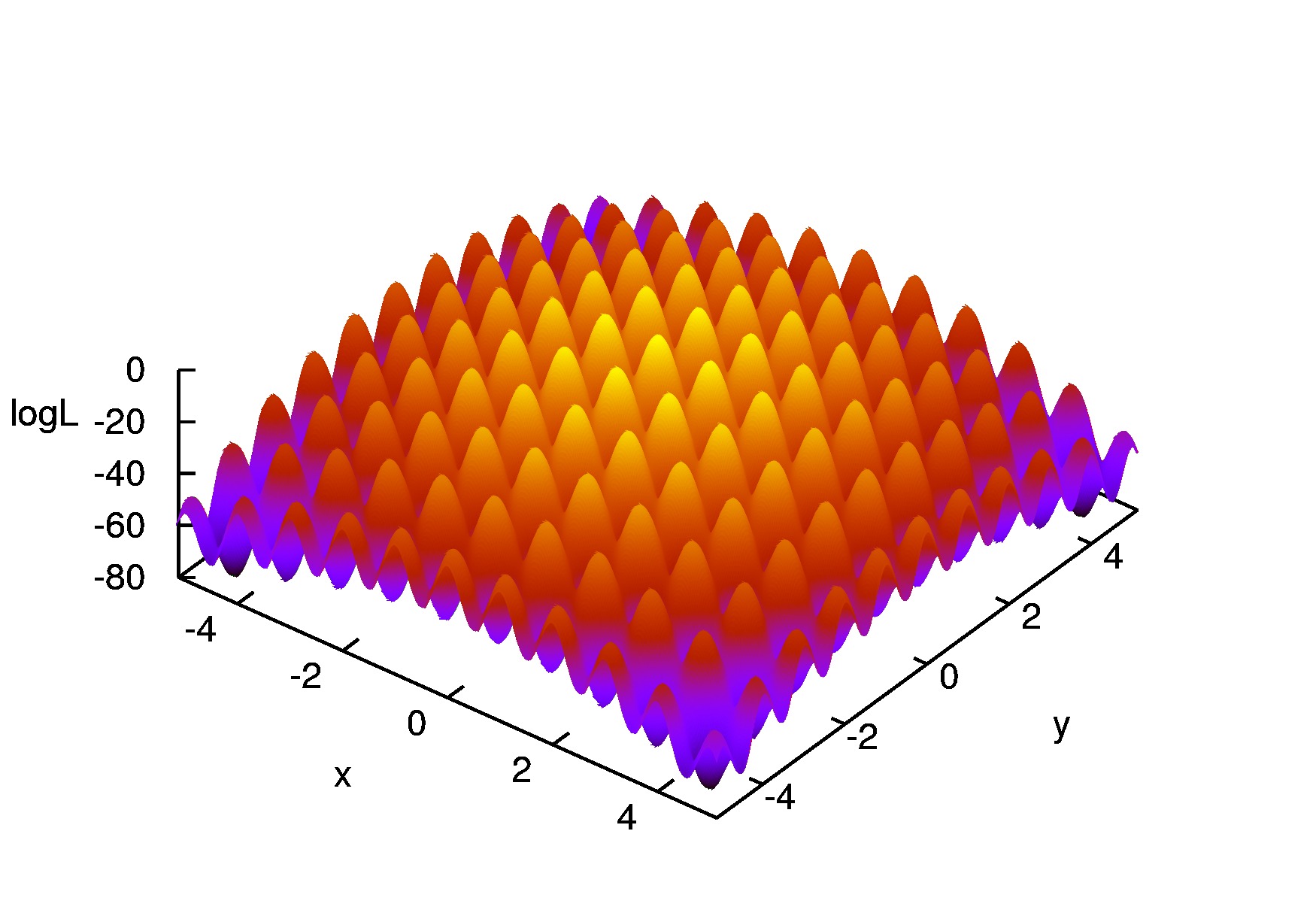}
\includegraphics[width=0.5\columnwidth]{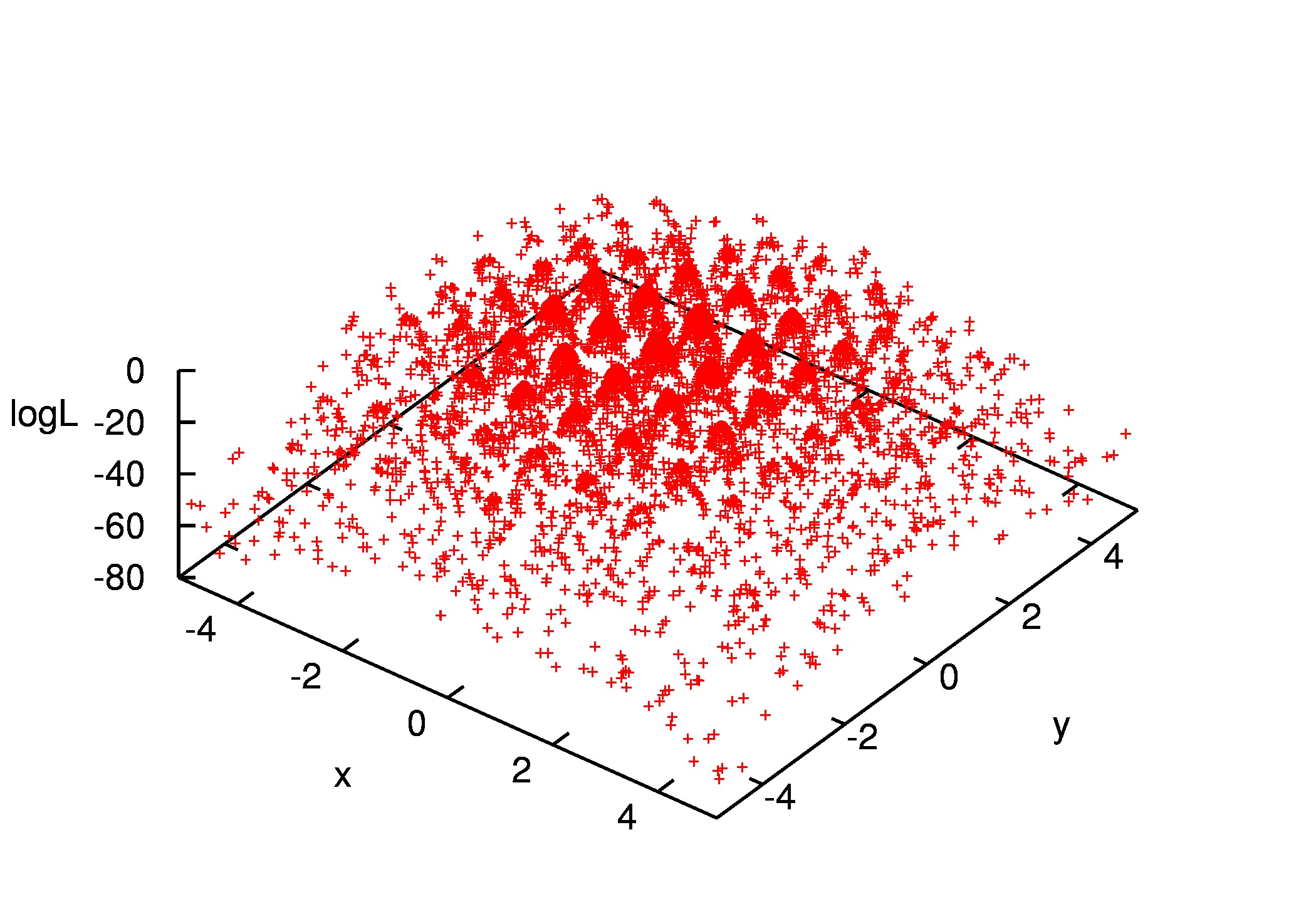}
\caption{Left panel: 2D Rastrigin function with the z-axis being $\log\mathcal{L}(\boldsymbol \Theta)$. Right panel: Samples obtained by running GMC algorithm on 2D Rastrigin function.}
\label{fig:rastrigin}
\end{figure}

Rastrigin function is defined as follows:
\begin{equation} 
f(\boldsymbol \Theta) = 10n + \sum_{\rm i = 1}^{n} \left[ \theta_{\rm i}^2 - 10 \cos(2 \pi \theta_{\rm i}) \right],
\end{equation}
where $\theta_{\rm i} \in [-5.12, 5.12]$. This functions has the global minimum at $\theta_{\rm i} = 0, \forall i$ where $f(\boldsymbol \Theta) = 0$. Fig.~\ref{fig:rastrigin} (left panel) shows a plot of the 2D Rastrigin function with the z-axis being $\log\mathcal{L}(\boldsymbol \Theta) = -f(\boldsymbol \Theta)$. GMC algorithm was run on this problem by assuming uniform priors $\mathcal{U}(-5.12, 5.12)$ on parameters $\theta_{\rm i}$. The algorithm took $215,916$ likelihood evaluations and the resultant samples are plotted in Fig.~\ref{fig:rastrigin} (right panel).

\subsection{Rosenbrock Function}

\begin{figure}
\includegraphics[width=0.5\columnwidth]{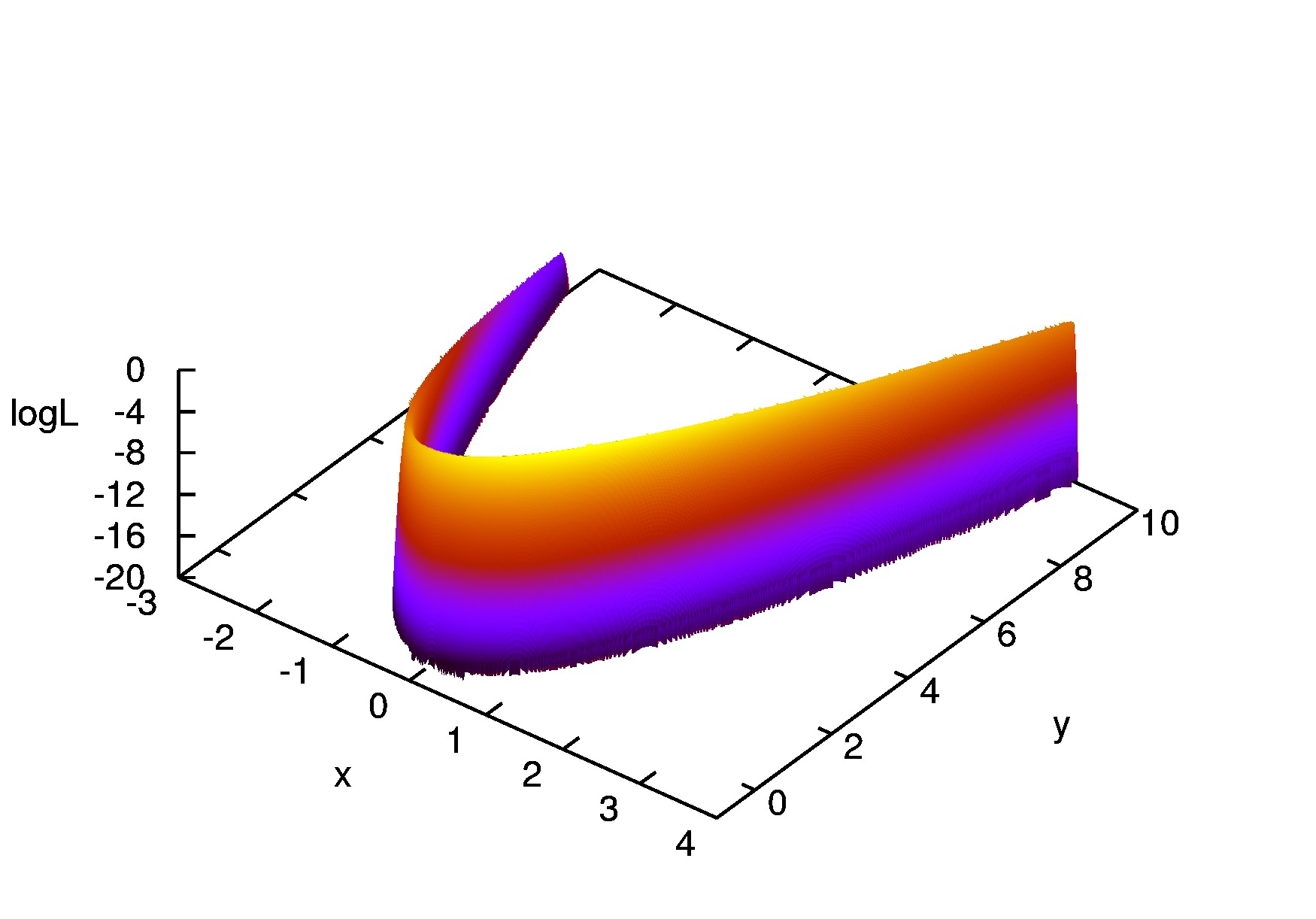}
\includegraphics[width=0.5\columnwidth]{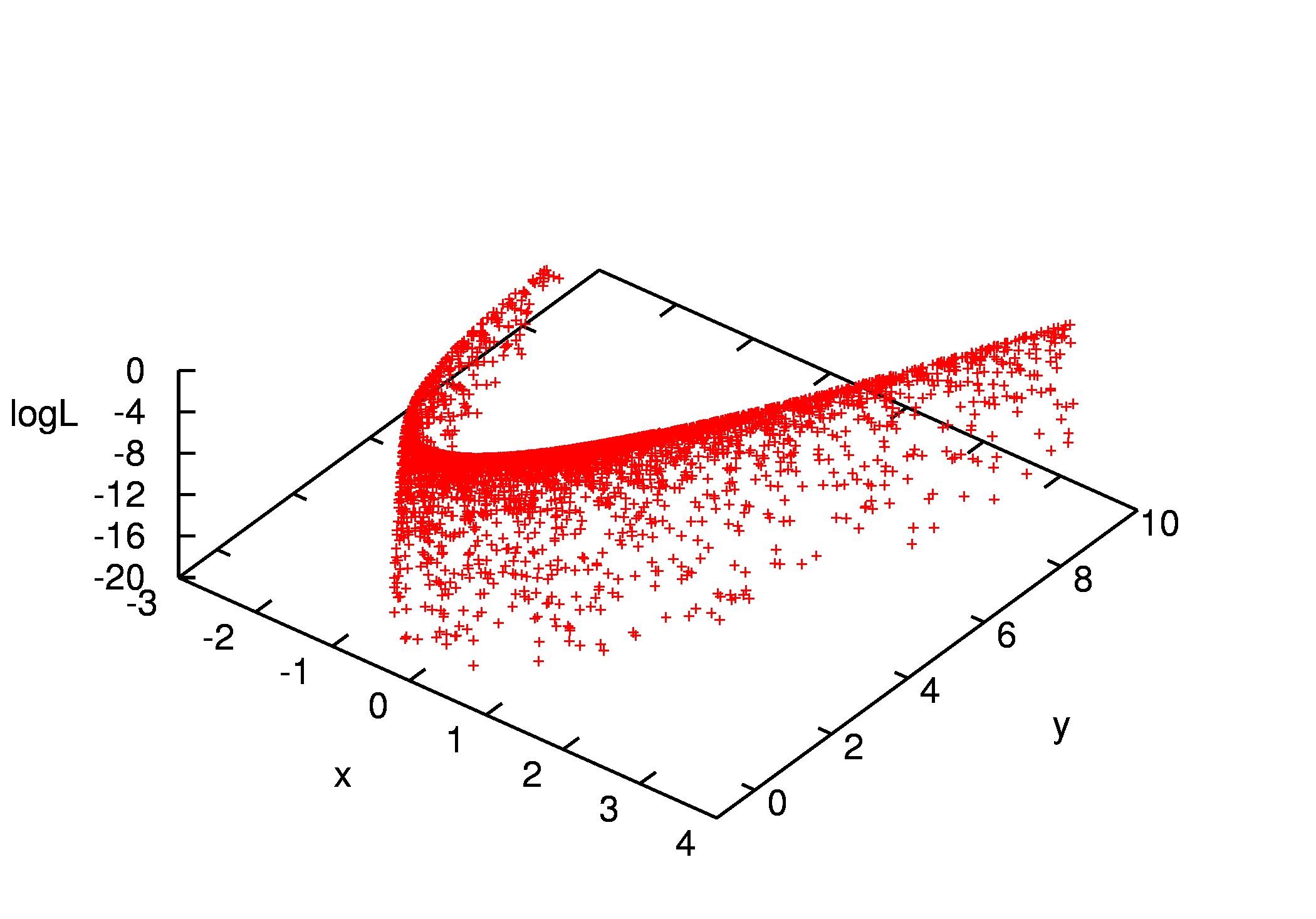}
\caption{Left panel: 2D Rosenbrock function with the z-axis being $\log\mathcal{L}(\boldsymbol \Theta)$. Right panel: Samples obtained by running GMC algorithm on 2D Rosenbrock function.}
\label{fig:rosenbrock}
\end{figure}

Rosenbrock function is defined as follows:
\begin{equation} 
f(\boldsymbol \Theta) = \sum_{\rm i = 1}^{n - 1} \left[ (1-\theta_{\rm i})^2 + 100(\theta_{\rm i+1} - \theta_{\rm i}^2)^2 \right],
\end{equation}
It has the global minimum at $(\theta_{\rm 1},\theta_{\rm 2},\cdots,\theta_{\rm n}) = (-1,1,\cdots,1)$ where $f(\boldsymbol \Theta) = 0$. Fig.~\ref{fig:rosenbrock} (left panel) shows a plot of the 2D Rastrigin function with the z-axis being $\log\mathcal{L}(\boldsymbol \Theta) = -f(\boldsymbol \Theta)$. Because of the presence of thin curving degeneracy, finding the global minimum for this problem is very challenging. GMC algorithm was run on this problem by assuming uniform priors $\mathcal{U}(-20,20)$ on parameters $\theta_{\rm i}$. The algorithm took $218,982$ likelihood evaluations and the resultant samples are plotted in Fig.~\ref{fig:rosenbrock} (right panel).

\section{Bayesian Object Detection}

We now consider the problem of detecting and characterizing discrete objects hidden in some background noise using Nested Sampling with GMC. We consider our data vector $\boldsymbol D$ to be pixel values in the image in which we want to search for these objects. Let us suppose that these objects are described by a template $s(\boldsymbol x; \boldsymbol \Theta)$, where $\boldsymbol \Theta$ denotes collectively the $(X,Y)$ position of the object, its amplitude $A$ and some measure $R$ of its spatial extent. In this example we assume spherical Gaussian shaped objects such that:
\begin{equation}
s(\boldsymbol x; \boldsymbol \Theta) = A \exp \left[ -\frac{(x-X)^2 + (y-Y)^2}{2R^2} \right]
\label{GaussObj}
\end{equation}
The contribution from each object is assumed to be additive. Therefore if there are $N_{obj}$ such objects present then:
\begin{equation}
{\boldsymbol D} = {\boldsymbol m} + \sum_{k=1}^{N_{obj}}s(\boldsymbol x; \boldsymbol \Theta_k),
\label{GaussObjData}
\end{equation}
where $\boldsymbol m$ denotes the generalized noise contribution to the data from background emission and instrumental noise and $s(\boldsymbol x; \boldsymbol \Theta_k)$ is the contribution to signal from $k^{th}$ discrete object. We therefore need to estimate the values of unknown parameters $(N_{obj}, \boldsymbol \Theta_1, \boldsymbol \Theta_2, \cdots, \boldsymbol \Theta_{N_{obj}})$.

\begin{figure}
\includegraphics[width=0.6\columnwidth]{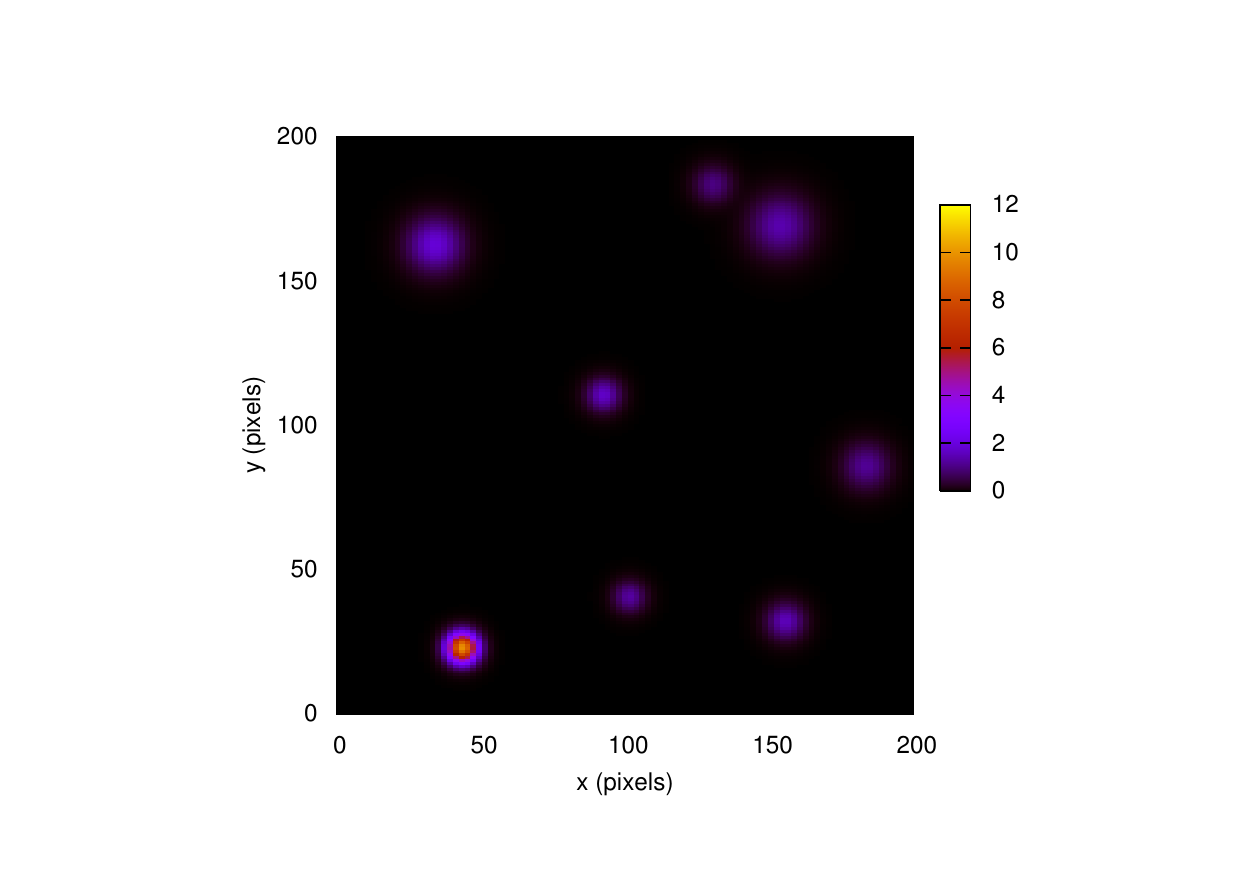}\hspace{-1.5cm}
\includegraphics[width=0.6\columnwidth]{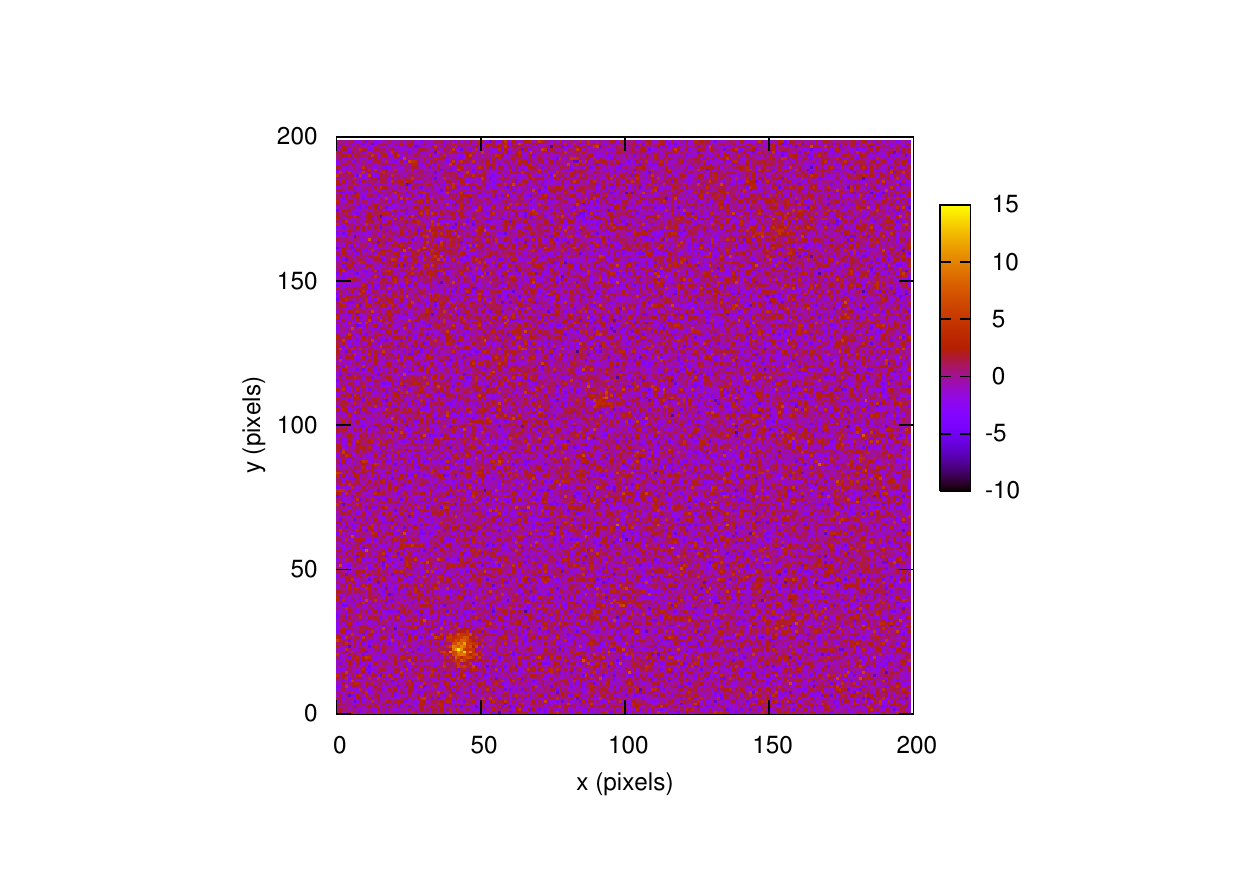}
\caption{The $200\times200$-pixel test image (left-hand panel) contains 8 Gaussian objects of varying widths and amplitudes; the parameters $X_k$, $Y_k$, $A_k$ and $R_k$ for each object are listed in Tab.~\ref{tab:ObjDetect}. Right-hand panel shows the corresponding data map with independent Gaussian noise added with an rms of 2 units.}
\label{fig:ObjDetect-Data}
\end{figure}

In order to test GMC on this problem, we simulated 8 objects in $200\times200$ pixel image, with template given in Eq.\eqref{GaussObj} and parameters listed in Tab.~\ref{tab:ObjDetect}. Final image is then created by adding independent Gaussian pixel noise with rms 2 units. The underlying model and simulated data are shown in Fig.~\ref{fig:ObjDetect-Data}.

One would ideally like to infer all the unknown parameters $(N_{obj}, \boldsymbol \Theta_1, \boldsymbol \Theta_2, \cdots, \boldsymbol \Theta_{N_{obj}})$ simultaneously from the data. This however requires any sampling based approach to move between spaces of different dimensionality as the length of the parameter vector depends on the unknown value of $N_{obj}$. Such techniques are discussed in \cite{Hobson03}. Nevertheless, due to this additional complexity of variable dimensionality, these techniques are generally extremely computationally intensive.

\begin{table}
\begin{tabular}{crrrrrrrr}
\hline
&  \multicolumn{4}{c}{True Parameter Values} & \multicolumn{4}{c}{Inferred Parameter Values}\\
Object & \myalign{c}{$X$} & \myalign{c}{$Y$} & \myalign{c}{$A$} & \myalign{c}{$R$} & \myalign{c}{$X$}  & \myalign{c}{$Y$} & \myalign{c}{$A$} & \myalign{c}{$R$} \\
\hline
1 &  43.7 &  22.9 & 10.5 & 3.3 & 43.4 $\pm$ 0.1 & 22.8 $\pm$ 0.1 & 11.0 $\pm$ 0.4 & 3.3 $\pm$ 0.1 \\
2 & 101.6 &  40.6 &  1.4 & 3.4 & 101.7 $\pm$ 1.0 & 41.2 $\pm$ 0.9 & 1.4 $\pm$ 0.3 & 4.1 $\pm$ 0.6 \\
3 &  92.6 & 110.6 &  1.8 & 3.7 & 92.0 $\pm$ 1.3 & 110.3 $\pm$ 1.2 & 1.4 $\pm$ 0.3 & 4.7 $\pm$ 0.9 \\
4 & 183.6 &  85.9 &  1.2 & 5.1 & 183.9 $\pm$ 1.6 & 87.0 $\pm$ 1.1 & 1.3 $\pm$ 0.3 & 4.8 $\pm$ 0.9 \\
5 &  34.1 & 162.5 &  1.9 & 6.0 & 34.0 $\pm$ 0.7 & 163.5 $\pm$ 0.8 & 2.0 $\pm$ 0.3 & 5.4 $\pm$ 0.5 \\
6 & 153.9 & 169.2 &  1.1 & 6.6 & 152.7 $\pm$ 1.1 & 170.5 $\pm$ 1.2 & 1.5 $\pm$ 0.2 & 6.1 $\pm$ 0.5 \\
7 & 155.5 &  32.1 &  1.5 & 4.1 & 157.2 $\pm$ 1.6 & 30.4 $\pm$ 1.3 & 1.2 $\pm$ 0.3 & 4.7 $\pm$ 0.9 \\
8 & 130.6 & 183.5 &  1.6 & 4.1 & 129.3 $\pm$ 1.1 & 183.2 $\pm$ 1.3 & 1.4 $\pm$ 0.3 & 5.0 $\pm$ 0.8 \\
\hline
\end{tabular}
\caption{True and inferred parameter values (with GMC) for $X_{k}$, $Y_{k}$, $A_{k}$ and $R_{k}$ $(k=1,...,8)$ defining the Gaussian shaped objects in Fig.~\ref{fig:ObjDetect-Data}.}
\label{tab:ObjDetect}
\end{table}

\begin{figure}
\includegraphics[width=0.3\columnwidth, angle=-90]{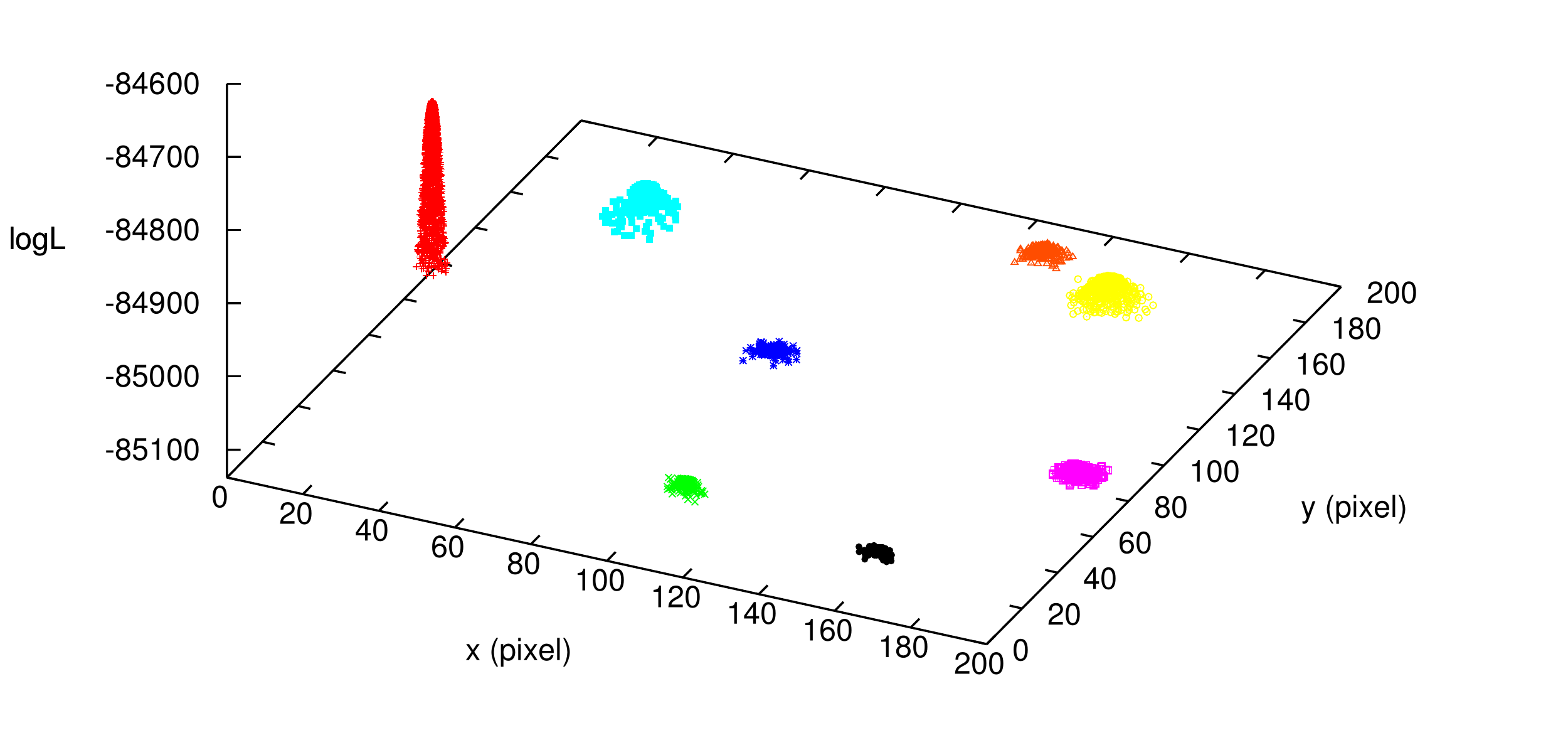}
\caption{The set of live points, projected into the (X, Y)-subspace, at each successive likelihood level in the nested sampling in the analysis of the data map in Fig.~\ref{fig:ObjDetect-Data} (right-hand panel) using GMC.}
\label{fig:ObjDetect-GMC}
\end{figure}

A similar problem was analysed in \cite{Hobson03} with a single source model and therefore the parameter space under consideration is $\boldsymbol \Theta = \left(X, Y, A, R\right)$ which is four-dimensional and fixed. This doesn't restrict us to detect only one object as the four-dimensional posterior distribution will have numerous modes, each one corresponding to the location of one of the real or spurious objects in the data. Due to high multi-modality, this represents a very difficult problem for traditional MCMC methods. \cite{feroz08} adopted the single source model for analysing this problem with the \texttt{MultiNest} implementation of Nested Sampling and showed that all 8 objects can be found very efficiently. Results of analysing this problem with GMC implementation of Nested Sampling, with a single source model are shown in Fig.~\ref{fig:ObjDetect-GMC} in which we plot the `live' points, projected into the $(X, Y)$-subspace, at each successive likelihood level in the Nested Sampling algorithm (above an arbitrary base level). Inferred parameter values for each object are listed in Tab.~\ref{tab:ObjDetect}. It can be clearly seen that all 8 objects have been identified. The run time with GMC was slightly higher than \texttt{MultiNest} but still orders of magnitude lower than MCMC approach used in \cite{Hobson03}.

\begin{theacknowledgments}
We would like to thank Mike Hobson and Steve Gull for very useful discussions on Nested Sampling.
\end{theacknowledgments}

\bibliographystyle{aipproc}   

\bibliography{FerozMaxEnt2012}

\begin{thebibliography}{7}
\expandafter\ifx\csname natexlab\endcsname\relax\def\natexlab#1{#1}\fi
\providecommand{\enquote}[1]{``#1''}
\expandafter\ifx\csname url\endcsname\relax
  \def\url#1{\texttt{#1}}\fi
\expandafter\ifx\csname urlprefix\endcsname\relax\def\urlprefix{URL }\fi
\providecommand{\eprint}[2][]{\url{#2}}

\bibitem[{Skilling}(2004)]{skilling04}
J.~{Skilling}, \enquote{{Nested Sampling},} in \emph{American Institute of
  Physics Conference Series}, edited by R.~{Fischer}, R.~{Preuss}, and U.~V.
  {Toussaint}, 2004, pp. 395--405,
  \urlprefix\url{http://www.inference.phy.cam.ac.uk/bayesys/}.

\bibitem[Sivia and Skilling(2006)]{sivia}
D.~Sivia, and J.~Skilling, \emph{Data Analysis A Bayesian Tutorial}, Oxford
  University Press, 2006.

\bibitem[{Feroz} and {Hobson}(2008)]{feroz08}
F.~{Feroz}, and M.~P. {Hobson}, \emph{\mnras} \textbf{384}, 449--463 (2008),
  \eprint{arXiv:0704.3704}.

\bibitem[{Feroz} et~al.(2009)]{multinest}
F.~{Feroz}, M.~P. {Hobson}, and M.~{Bridges}, \emph{\mnras} \textbf{398},
  1601--1614 (2009), \eprint{arXiv:0809.3437}.

\bibitem[{Skilling}(2012)]{2012AIPC.1443..145S}
J.~{Skilling}, \enquote{{Bayesian computation in big spaces-nested sampling and
  Galilean Monte Carlo},} in \emph{American Institute of Physics Conference
  Series}, edited by P.~{Goyal}, A.~{Giffin}, K.~H. {Knuth}, and E.~{Vrscay},
  2012, vol. 1443 of \emph{American Institute of Physics Conference Series},
  pp. 145--156.

\bibitem[Li et~al.(2009)]{DBLP:journals/amc/LiPAZG09}
Y.~Li, V.~A. Protopopescu, N.~Arnold, X.~Zhang, and A.~Gorin, \emph{Applied
  Mathematics and Computation} \textbf{212}, 216--228 (2009).

\bibitem[{Hobson} and {McLachlan}(2003)]{Hobson03}
M.~P. {Hobson}, and C.~{McLachlan}, \emph{\mnras} \textbf{338}, 765--784
  (2003), \eprint{astro-ph/0204457}.

\end{thebibliography}

\end{document}